\documentclass[twocolumn,prl,showpacs]{revtex4}

\usepackage{graphicx}
\usepackage{amssymb,amsfonts,amsmath}
\usepackage[center]{subfigure}

\begin{document}

\title{Emergence of heterogeneity and political organization in information exchange networks}
\author{Nicholas Guttenberg}
\affiliation{James Franck Institute,
University of Chicago, Gordon Center for Integrative Science, 929 E 57th Street, Chicago IL 60637}
\author{Nigel Goldenfeld}
\affiliation{Department of Physics, University of Illinois at
Urbana-Champaign, 1110 West Green Street, Urbana, Illinois, 61801-3080.}

\begin{abstract}
We present a simple model of the emergence of the division of labor and
the development of a system of resource subsidy from an agent-based
model of directed resource production with variable degrees of trust
between the agents. The model has three distinct phases, corresponding
to different forms of societal organization: disconnected (independent
agents), homogeneous cooperative (collective state), and inhomogeneous
cooperative (collective state with a leader).  Our results indicate
that such levels of organization arise generically as a collective
effect from interacting agent dynamics, and may have applications in a
variety of systems including social insects and microbial communities.
\end{abstract}


\maketitle

\section{Introduction}

To understand systems of social and political organization, it is
tempting to begin by trying to understand the individuals that form
them. This approach quickly runs into a problem---the behavior of
individuals is very hard to predict. The behavior of any given
individual depends upon a large number of factors: their culture, their
experiences to date, their genetics, the events they are currently
experiencing, their education, their economic status, and so on. It
seems as if understanding the behavior of a group is an impossible goal
if predicting the behavior a single person is so difficult. However,
models of group behavior through agent-based modelling\cite{BONA02}
have been reasonably successful despite this, reproducing generic
properties of the dynamics of crowds, mobs, and
riots\cite{GRAN78,WOHL83,HELB02}; collective opinion
formation\cite{ROBI01, HUET08, NARD08}; the structure of social
groups\cite{ZACH77,skyrms2000dms,JIN01,STRO02,GEAR08,PACH06,SHOH97,BOEL06};
and financial markets\cite{KIRO08}. When large numbers of people
interact, there exists the possibility for the emergence of collective
effects which---surprisingly---are insensitive to the details of the
elements which comprise them.  When this occurs, the interactions
between agents overwhelm their individual dynamics; and although their
may well be many factors difficult to model for each individual's
behavior, the interactions are frequently easier to specify,
characterize and model.

The purpose of this paper is to understand the factors at work in
setting up and maintaining the large scale structure of societies from
the point of view of an abstract model. Other models\cite{GAND06} have
analyzed the stability and transitions of an established form of social
order. In this paper, we will instead seek to explain how social order
emerges from an unstructured state due to collective interactions
between individual agents. This must take into account that the
connections between individuals may change, leading to a situation in
which one has an active network\cite{GROS08}.

The emergence of networks of preferred interactions between agents has
been observed in \cite{ZIMM05,EGUI05}. The resultant structure of
agents is heterogeneous---a state emerges in which some subset of the
population (the leaders) extracts maximal benefit. There is however no
explicit flow of information from the leaders to the other agents. We
posit that the structure of information exchange in the system is a key
element to the form of political organization it possesses. We would
like to differentiate between the agent with the greatest payoff and
the agent whose decisions hold maximal weight in influencing the
decisions of others. In our model, we observe the development of a
division of labor from simple selfish behavior and communication
between the members of the system. The role of active information is
central to achieving this heterogeneous population structure. This
mechanism is not unique as far as ways in which the division of labor
might emerge\cite{romer1987growth,robinson1992regulation}. Any system that encourages specialization and provides
some way for the proceeds of labor to be redistributed may very well
produce division of labor, and there are a number of proposals for how
this might come about.

The mechanism studied in the present paper is differentiated from
earlier work because the method of redistribution (information
exchange) has the additional consequence that networks of behavioral
control can emerge from the population. They are not mandatory
consequences of the dynamics, but only arise under certain conditions
specified by parameters in the model. Thus our model does not function
as a zero-sum game in which there is exchange of a variety of
resources.  Information exchange has been studied in various other
models. However, in such models it is usually a passive variable, for
instance in voting and opinion formation models.

The role of active information---information used to make a decision
with either positive or negative consequence---is less well known. In
\cite{ZIMM02}, active information played the role of a diffusive field
in a spatial prisoner's dilemma model, and in \cite{COUZ05},
information was given to a subset of members of a swarm to see how
informed decisions would propagate to determine the swarm direction. In
these cases, the agents had no way of evaluating the quality of the
information they received---whether it had in the past led to a good or
bad decision. This dynamics leads to information acting primarily as a
homogenizing agent: it determines the average behavior in
\cite{ZIMM02}, and directs the average swarm direction in
\cite{COUZ05}.  On the other hand, in our model, each agent determines
the optimal degree of trust to place in information received from
another. This `trust', in other contexts such as a political system or
organizational structure, could be any way in which
control over an agent's behavior is surrendered to one or more external
agents. By giving each agent the ability to tune its trust in the other
members of the system, it is possible for clusters to form in which the
members of the cluster have voluntarily given over the reins of their
decision making to a leader of their choice.

This organization, in its simplest form, arises from uniform
information exchange between the individuals in the system, resulting
in a homogeneous, shared information pool. This corresponds to communal
decision-making by majority vote. In a system in which different agents
are better or worse at making decisions, one would expect the emergence
of a system of weighting by reputation, simply as a tool to optimize
the decision-making process. If, however, resources can be allocated
towards making better decisions, it becomes possible for a subset of
the individuals to specialize in being an information source. At this
point,  the majority of agents in the system will be following
instructions provided by a minority of agents, without a significant
information flow in the reverse direction. These two
phases---unstructured and structured respectively---are distinct forms
of political organization, and which is achieved depends on the costs
and benefits associated with information generation.

\begin{figure}[t]
\includegraphics[width=\columnwidth,angle=0]{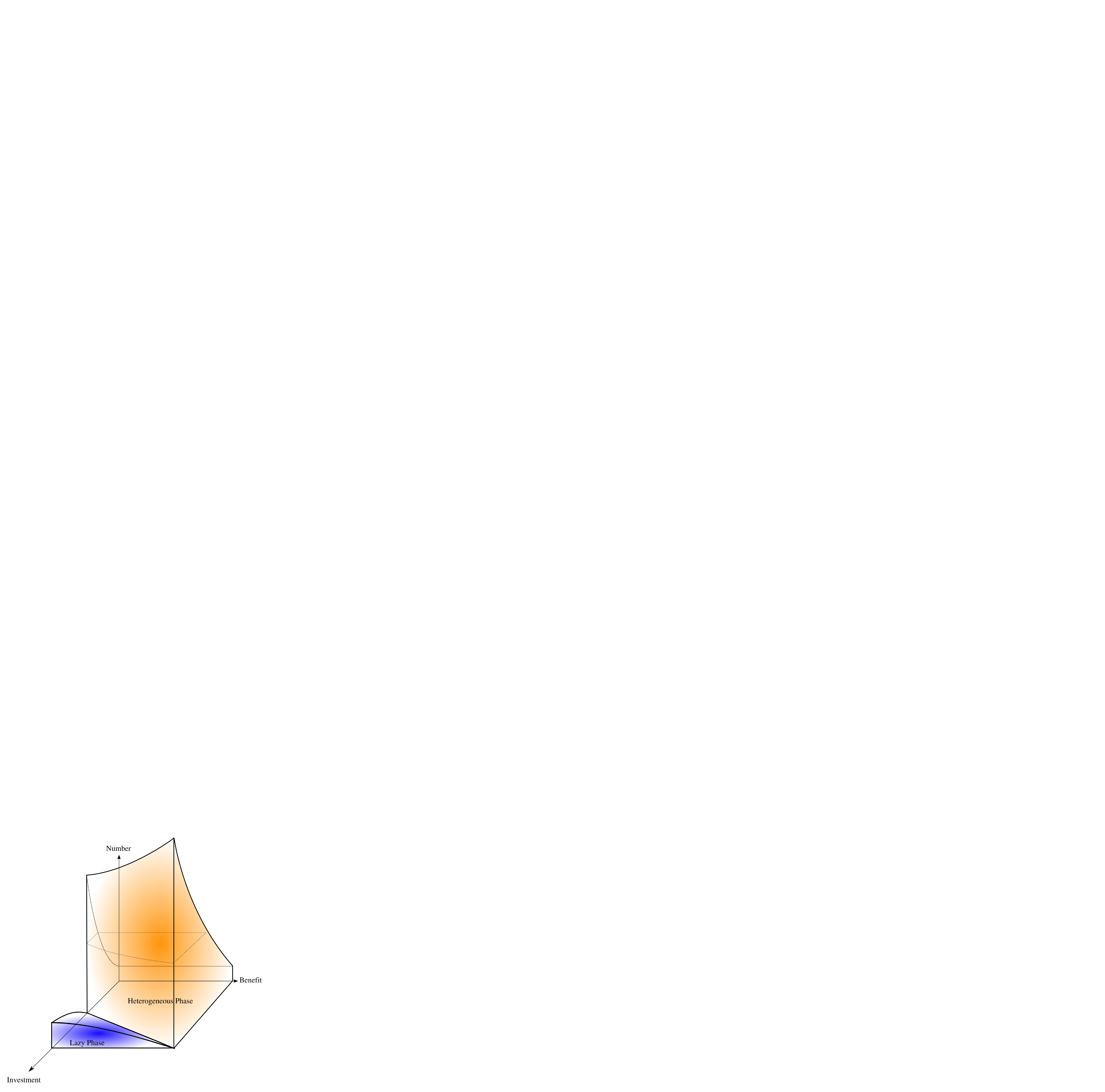}
\caption{Schematic phase diagram for our model. The Investment axis is
the degree that a large initial investment of resources is needed to
see an improvement in accuracy: this corresponds to the nonlinearity
$\alpha$ in the model. The benefit axis is the total difference in
accuracy between random guessing and perfect knowledge, which
corresponds to the variable $O$ in our model.  In the Lazy Phase,
random guessing is the optimal behavior. In the Heterogeneous Phase, a
subset of agents dedicate their resources to thinking whereas the rest
of the agents dedicate their resources to working (division of labor).
In the Homogeneous Phase), all the agents dedicate the same
non-zero amount of resources to thinking.}
\label{Summary_phase_diagram}
\end{figure}

A requirement for stability in the structured phase is that the agents
which are acting as an information source must either gain from
producing information or lose if they fail to produce information, as
they dedicate their own resources into providing this information. In
modern governments, systems of taxation subsidize the decision-makers,
but the emergence of such structures is difficult without a
heterogeneous system already being in place. Our results show that in
certain circumstances, the decision-making structure of a population
may become heterogeneous even without the inclusion of subsidies or
resource exchange, due to a collective effect where the refusal to
generate information by the majority of the population forces the
agents that are the last to act to take on the decision-making role
simply to preserve their own benefit. From this phase, the introduction
of a resource subsidy would improve the efficiency of the system, and
could be done in a continuous manner.  A schematic phase diagram that
qualitatively exhibits the nature of the phases and transitions between
them is illustrated in Fig.~(\ref{Summary_phase_diagram}).

\section{Model}

We propose the following model to capture the dynamics of information
exchange. The system consists of a set of agents, each which can choose
to distribute resources to any other agents. In addition, each agent
chooses to allocate its time between producing resources or producing
information about the environmental state (\lq thinking'). Whether or not
resource production is successful depends on the accuracy of the
agent's guess as to the current nature of its environment, which is
randomly in one of $O$ possible states. If the agent guesses the
environmental state correctly, it produces a number of resources
proportional to the fraction of its time it allocated to production.
Furthermore, an agent can look to see what other agents are guessing in
order to determine its own guess.

We assume in this first part that each agent has a number of degrees of
freedom (how to combine information from other agents, how many
resources to distribute to other agents, and how much time to allocate
towards producing resources) which are adjusted in order to maximize
its average score. The immediate consequence of this is that we may
determine trivially what the trust network should be, and thus
determine our trust-network order parameters in terms of the
distribution of '\lq thinking' values---the agent with the highest thinking
value will have the most trust directed at it, and if the thinking
values are distributed homogeneously then trust will also be
distributed homogeneously. This treatment neglects dynamical effects
and fluctuations. Later, we will  analyze the effect of fluctuations
and dynamics on the stability of the various phases.

The base accuracy---that due to the agent's own production of
information, is a nonlinear function of the fraction of time dedicated
towards information production $T$. A successful guess then produces
one resource per unit time spent on resource generation. This results in a total production of
$1-T$ resources from a successful guess, or zero from a failed guess. If we take the average
performance over many such trials, then we can derive a score function that the system may
try to optimize.

We must now determine how the accuracy depends on the amount of time spent upon `thinking'.
The choice of functional form must satisfy a number of
constraints. The accuracy should monotonically increase with the
fraction of time dedicated towards it. Additionally, it is bounded
above by $1$ and below by $1/O$ (the accuracy of a random guess). Given
these constraints, we may choose any function of the form
$A=(1/O+(1-1/O)f(T))$ where $f(T)$ is a monotonically increasing
function that maps the interval $[0,1]$ to itself.

The key character of our choice of function will be the range of values
of the other parameters for which the score function has a local maximum in the interval $[0,1]$.
The point at which this maximum appears or disappears will control part of the resultant phase diagram.
If $f(T)$ is monotonically increasing, then the larger $O$ is, the more likely there is for there to be a maximum,
and the less concave up $f$ is, the more likely there is to be a maximum. This can be seen by calculating the concavity
of the total score function $S$ at the location of its extrenum in terms of an arbitrary $f$:

\begin{equation}
\textrm{sgn}(S^{''}) = \textrm{sgn}( (1-x)^2 f^{''} - 2 (f+\frac{1}{O-1}))
\end{equation}

Consequently, the specific details of $f(T)$ should not strongly influence the results. Its local derivatives in the vicinity
of the extrenum of $S$ are the only relevant properties. If $f(T)$ is concave up, then specialization is favored. If concave down,
then there are diminishing returns and even an infinitesimal amount of time dedicated towards producing information will be beneficial. While we could in principle combine an arbitrary number of concave up and concave down regions in order to create a series of optima in $S$, it is hard to justify that arbitrary
complication.

A simple choice of function that allows us to smoothly
vary between concave up and concave down behavior with a single
parameter is $A = (1/O + (1-1/O)T^\alpha)$, where $\alpha$ is a
parameter of the model controlling the disposition of the social problem towards specialization or generalization.
If $\alpha>1$ then the function is concave up, and specialization is favored. With this
basis, we can discuss a number of possible system configurations and
evaluate their average score for optimal choices of $T$.

While one could argue that we have put in the possibility of the existence of an optimum value of $T$ by hand, it is an allowed possibility given the most arbitrary choice of $f(T)$. We are then exploring the consequences to the phase diagram of political organization that result from the existence of this optimum choice, rather than saying that we have shown that fundamentally that optimum must exist in real social systems. It is clear that in many cases such as social insects there is such an optimum, because in those systems specialization is favored over generalization.

\subsection{Disconnected, Homogeneous Phase}

In the case that no agent in the system uses information from any other
agent, there is an optimal value of $T$ to maximize an agent's score.
The average score in this phase $\bar{S_{DH}}$ is perforce independent
of $N$.

\begin{equation}
\bar{S_{DH}}(T) = \frac{1}{O}(1-T)(1+(O-1)T^{\alpha})
\end{equation}
\noindent
The optimal value of $T$ satisfies:

\begin{equation}
T^{\alpha-1}(\alpha-(\alpha+1)T) = \frac{1}{O-1}
\end{equation}

\begin{figure}[t]
\includegraphics[width=\columnwidth,angle=0]{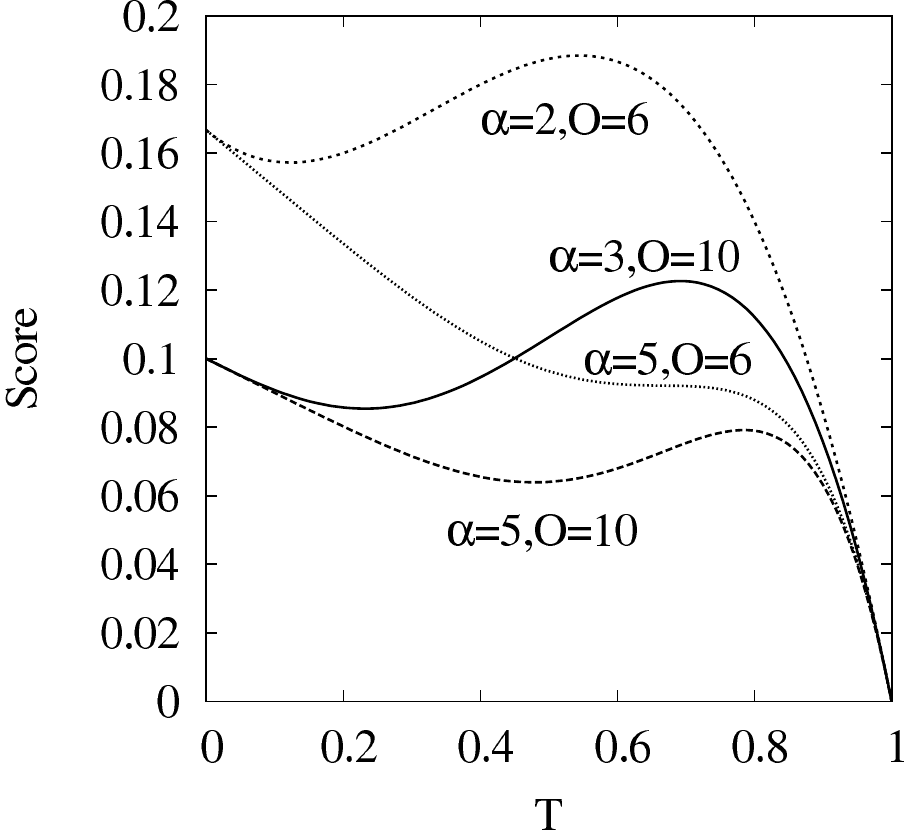}
\caption{Score functions in the Isolated Phase as a function of
thinking time $T$, for four values of $O$ and $\alpha$.  In the
Isolated Phase, each agent does not receive information from other
agents. } \label{ScoreFunc1}
\end{figure}

If $\alpha=1$, then this value of $T$ is always less than zero, so
$T=0$ is the optimal choice. At larger values of $\alpha$ a local
maximum appears in the curve at a finite value of $O$, and then becomes
a global maximum as $O$ increases. The value of $O$ at which the
maximum value of the score is equal to the value at $T=0$ is
$O=1+\alpha^\alpha / (\alpha-1)^{\alpha-1}$. In the limit of large
$\alpha$, this becomes $O\approx e\alpha+(1-e/2)$. So in effect, for
values of $\alpha>1$ (representing a nonlinear reward for dedicating
resources to \lq thinking') there is a first order transition between a
\lq guessing' phase and a \lq thinking' phase, where the more options there
are, the more valuable a resource spent on \lq thinking' is. The larger
$\alpha$ is, the larger $O$ must be for a non-zero thinking phase to be
optimal. The score function for various values of $\alpha$ and $O$ is
plotted in Fig.~\ref{ScoreFunc1}.

An additional consideration is the effect of fluctuations on this
phase. If each agent may only specify their actual thinking value to
within some standard deviation, then the resulting average score is
lower than if fluctuations had been absent. Near the limits of the
range of the thinking variable fluctuations are constrained such that
they may not take it outside of the range.  For fluctuations of
magnitude $\sigma$ around an optimal value of $T$, we expect that the
average score will change by:

\begin{equation}
\Delta S = \sigma^2 \frac{d^2 S}{dT^2} = \sigma^2 \alpha T^{\alpha-2} ( (\alpha-1) - (\alpha+1)T ) \frac{O-1}{O}
\end{equation}

For fixed $\alpha$, as $O$ becomes large the optimal value of $T$
approaches $\alpha/(\alpha+1)$ and so the decrease in the score
approaches:

\begin{equation}
\Delta S = -\alpha \left(\frac{\alpha}{\alpha+1}\right)^{\alpha-2} \sigma^2
\end{equation}

\noindent
When the optimal solution is $T=0$, however, the first derivative is
non-zero and so fluctuations have a linear effect. The effect of this
is that $\Delta S = \sigma \frac{d S}{d T} / \sqrt{\pi}$ assuming
Gaussian fluctuations. The slope of the score function around $T=0$ is:

\begin{equation}
\frac{d \bar{S_{DH}}}{dT} = -1/O
\end{equation}
\noindent
So we expect $\Delta S = -\sigma/(O\sqrt{\pi})$ to be the leading
effect at this point. The consequence of this is that sufficiently
large fluctuations will favor the $T=0$ phase.

\subsection{Connected, Homogeneous Phase}

If communication between agents is permitted, but no resource
reallocation takes place, then the resulting accuracy is higher than
any of the individual accuracies in the system (so this phase is always
favored over the disconnected phase for permitted values of $T$).

For $O=2$, the effective accuracy can be solved for in the large $N$
limit. If the initial accuracy is $A$, then the total number of agents
that pick the correct option $C_0$ is $C_0 = \sum_i^N \eta_i$ where
$\eta_i$ is either $1$ (with chance $A$) or $0$ (with chance $1-A$). In
the large $N$ limit, $C_0$ is described by a Gaussian distribution with
mean $NA$ and standard deviation $A(1-A)\sqrt{N}$.

The probability that the system picks the correct option is thus the
probability that $C_0 > N/2$. As the range of permitted values is not
infinite, care must be taken to compute the correct normalization
factor. So:
\begin{equation}
A_{eff} = \frac{\int_{N/2}^{N} \exp(-(\frac{x-AN}{2NA(1-A)})^2)}{\int_{0}^{N} \exp(-(\frac{x-AN}{2NA(1-A)})^2)}
\end{equation}
\noindent
which evaluates to
\begin{equation}
A_{eff} = \frac{\mathrm{erf}(\frac{(1-A)\sqrt{N}}{\sqrt{2A(1-A)}}) - \mathrm{erf}(\frac{(1/2-A)\sqrt{N}}{\sqrt{2A(1-A)}})}{\mathrm{erf}(\frac{(1-A)\sqrt{N}}{\sqrt{2A(1-A)}})-\mathrm{erf}(-\frac{A\sqrt{N}}{\sqrt{2A(1-A)}})}
\end{equation}
\noindent
where $A(T) = (1+T^{\alpha})/2$ in this case.

\begin{figure}[t]
\includegraphics[width=\columnwidth,angle=0]{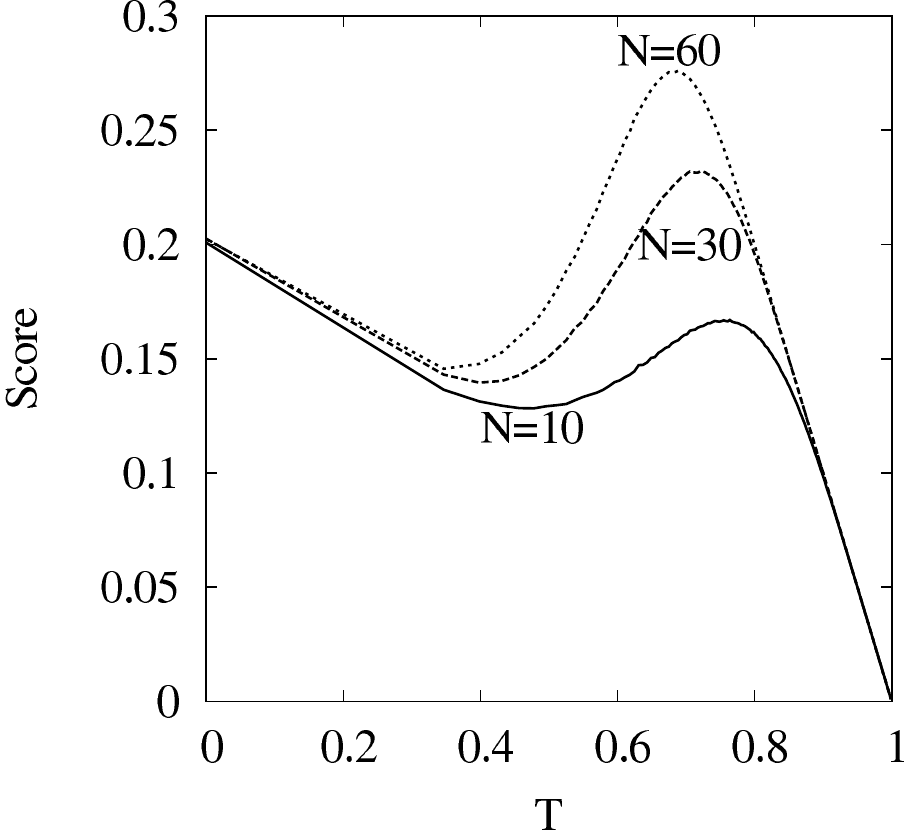}
\caption{Score functions in the Homogeneous Phase as a function of
thinking time $T$, for different numbers of agents $N$, with  $O=5$
and $\alpha=5$.  In the Homogeneous Phase all agents share information
and have the same parameters.} \label{ScoreFunc2}
\end{figure}

For $O=2$ and $\alpha=2$, benefits from a non-zero value of $T$ do not
appear until around $N>35$. Figure \ref{ScoreFunc2} shows the score
function for the homogeneous, connected phase compared with the
isolated phase.

The effects of fluctuations are less obvious in this case, because they
must be considered each agent independently, whereas this analysis is
done for all agents behaving in the same fashion. In the case of this
model, fluctuations may actually increase the effective score, as a
fluctuation to higher thinking rate in one agent benefits the guesses
of all other agents. Similarly, a decrease in thinking rate in one
agent will not significantly decrease his accuracy but may increase his
yield. This is a hint that this particular phase is unstable to an
inhomogeneous phase.

A rough estimate would suggest that when adding together the effects of
fluctuations on each of the individual agents, the effective size of
fluctuations is reduced from $\sigma$ to $\sigma^{\prime} =
\sigma/\sqrt{N}$. This has the consequence that the  connected,
homogeneous phase is less sensitive to fluctuations than the
disconnected phase.

\subsection{Connected, Inhomogeneous Phase}

If the agents become inhomogeneous and divide their labor between
thinking and working, then structures in which there is a directional
information flow become possible. Given perfect communication and no
fluctuations, the optimal configuration will be  that of a single agent
with high accuracy ($T$), and $N-1$ agents with minimum accuracy but
always picking the action of the \lq leader' agent. The average score for
this phase is simply:

\begin{equation}
\bar{S_{CI}} = \frac{(N-1)+(1-T)(1/O + (1-1/O)T^{\alpha}}{N}
\end{equation}

\noindent
This phase in static conditions scores far better than the homogeneous
phases, but it is very susceptible to fluctuations lowering the score,
compared to the connected homogeneous phase. The result of this is that
neither the pure homogeneous nor heterogeneous phases are realized. In
a fully-connected population with some form of noise, the system
produces a number of leaders $L$ which scales with the population size.

The inhomogeneous phase with a number of leaders can always have a
higher average score than the homogeneous phase, but is not generally
stable when the individual scores are examined. Each leader agent can
improve their score by decreasing the portion of resources they
dedicate to thinking to the optimal value for the disconnected phase.
When the disconnected phase optimal value is $T$ is greater than zero,
the inhomogeneous phase may still occur. This occurs for small
$\alpha$, large $O$, and small $N$. If $\alpha$ is too large, the
height of the secondary maximum is decreased below that of the $T=0$
score function maximum and a homogeneous $T=0$ phase occurs. If $N$ is
sufficiently large, the homogeneous connected phase with nonzero $T$
can outperform a phase consisting of a single \lq selfish leader'. So
there are first order phase transitions in the space of $O$, $\alpha$,
and $N$ between three phases: $T=0$, leader, and homogeneous connected
(or \lq communal' phase).

\subsection{Resource Subsidy}

We have so far shown that for certain values of the parameters, the
inhomogeneous \lq leader' phase is stable even without the leaders being
subsidized. The system has not maximized its resource production in
this phase---rather, the limit on resource production is set by the
cost to the leader agent, in that even though it might produce a large
amount of resources for others by changing its behavior, doing so would
decrease its own resource production.

If we allow agents to exchange resources as well as information, then
starting from the connected, inhomogeneous phase it is possible to
improve or keep constant the scores of all agents. If we have a phase
with a single leader agent, then for that agent to dedicate more than
the disconnected optimal fraction of resources to thinking, it must be
reimbursed by at least the same amount of resources as it loses to
increase the resources it spends on thinking. This resource cost may
then be absorbed by the remaining $N-1$ agents. In effect, the
criterion of selfish optimization becomes one of global optimization.
The globally optimal phase in the absence of fluctuations is that with
a single leader agent.

This need not be the case in general, as one may posit the existence of
cheaters: agents which do not give resources towards the subsidy but
still gain its benefits. A system with multiple types of resource or
multiple agendas, such as in \cite{TROF02} might also retain a more
detailed structure.

\begin{figure}[t]
\includegraphics[width=\columnwidth,angle=0]{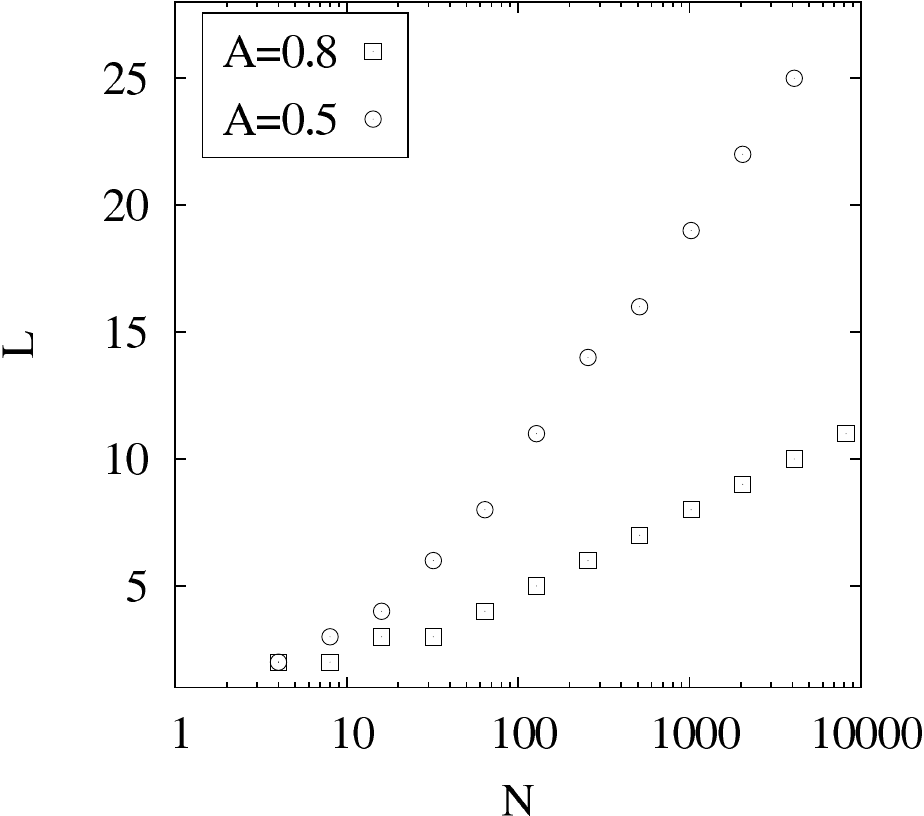}
\caption{Optimal number of leaders as a function of total number of
agents for a system with $O=5$ and two different fluctuation strengths.
In this case, the fluctuations are parameterized by the resultant
average accuracy $A$ of a leader with $T=1$. } \label{Leaders}
\end{figure}

When fluctuations are added, it becomes beneficial to have multiple
leaders in order to reduce the impact of fluctuations but retain the
benefit of increased efficiency. We use the connected, homogeneous
solution for $L$ agents to determine the accuracy of the remaining
$N-L$ given a known accuracy of the leaders. For simplicity, we will
assume that the leader agents have $T=1$, which the optimal choice
converges to as $N \gg L$. For a given level of fluctuations, each
leader will have an effective, adjusted accuracy. We evaluate the score
function numerically as a function of $L$ and find the location of the
maximum as a function of $N$. The results are plotted in
Fig.~\ref{Leaders}. At large $N$, the optimal number of leaders
approaches $L \propto \log(N)$.

For a spatially distributed system, or one in which there is not total
connectivity, it is expected that such effects will require a larger
number of leaders to cover the system extent. For example, in a
two-dimensional system in which agents can only communicate within a
radius $R$, a number of leaders proportional to $\sqrt{N}/\pi R^2$
would be expected to ensure total coverage.

\section{Applications}

The abstract model of emergent political systems that we have outlined
is capable of providing a framework in which to analyze real social
systems, and in this section we briefly indicate some examples. It is
important to emphasize that our model is not required to model all
situations in which division of labor occurs---a simpler model with
only a nonlinear benefit to specialization and some form of exchange of
services would be sufficient to enable division of labor to emerge.
Such a process would not need to involve information sharing as a core
element. On the other hand, our work shows that the emergence of the
leader phase (which corresponds to the occurrence of a division of
labor in other pictures) is primarily a consequence of the special
property of information when compared to other resources that, once
created, it can be duplicated with a much smaller additional cost than
the cost to first generate it. This process of information
amplification makes the leader phase described here distinct from other
scenarios that produce division of labor.

We must also be careful to understand the nature of the relevant
optimization being implicitly performed when considering a given
system. In human economic and political behavior, one considers that
each individual tries to maximize its personal benefit in the context
of the greater system. In other systems, such as foraging insects, the
net benefit to the colony as a whole is what is likely maximized---this
corresponds to the case where resources may be redistributed, which in
our model means that the leader phase is always optimal for all
parameter values.

Even with these caveats, there are several systems which could
potentially be understood in the context of our model: the behavior of
social and hierarchical insects compared to asocial
insects\cite{evans1966bps,richter2000swh,robinson1992rdl}, the distribution of
information in swarms\cite{seeley1999gdm}, and innovation-sharing in
unicellular organisms via horizontal gene transfer\cite{hecht2006cpt,
vetsigian2006cea}. All of these cases involve some piece of information
being discovered by a single individual---a randomly chosen one of a
set of similar individuals in the case of swarm behavior (corresponding
to the homogeneous phase), or via directed searching by a specialized
subset of the population, as is the case in some foraging insects (corresponding to the leader phase).
We will now briefly discuss each case.

Different species of insects are socialized to different degrees. On
one extreme, there are insects such as the solitary
wasps\cite{evans1966bps}, which do not share resources or information.
On the other extreme, eusocial insects such as bees, ants, and certain
kinds of wasps have highly structured communication channels and
vehicles of information discovery. Foragers and scouts use various
means to communicate the location of food supplies or nesting sites.
The distinction here seems to be that bees and ants reproduce centrally
via a queen, and so maximizing their interest corresponds to maximizing
the interest of the queen.  As a result, resources can be redistributed
freely, and so we expect the system to emerge in the leader phase - this is equivalent to a
system of resource subsidy as discussed earlier. `Trust', here, is embodied in the
genetically programmed behaviors of the individuals in following signals sent by other
insects.

In the case of honey bees, the various scouts return with information
about potential food locations, after which the swarm comes to a
unified decision about which site to pursue. The method of decision
making seems to be a weighted average\cite{seeley1999gdm}, similar to
what we use to model the decision making of our agents. Each scout has
a certain chance of finding the best site within a given
distance---even if they spend 100\% of their time searching, they have
a limited maximum accuracy. This corresponds to the fluctuating case in
our model, so, as the swarm size grows, we can predict that the optimal
number of scouts should scale logarithmically with the swarm size.

Microbial organisms\cite{ochman2000lgt,AnaBabic03142008} and even
multicellular eukaryotes\cite{hotopp2007wlg, gladyshev2008mhg,
keeling2008hgt, pace2008rht} have the ability to swap genetic material
and integrate it into their genomes via several pathways, mediated
often by mobile genetic elements such as viruses and plasmids. There is
cellular machinery associated with this process, which can be active or
inactive in a given cell. In microbes, the state in which such an
organism is receptive to external DNA is called genetic competence. The
regulatory network associated with competence has been shown to
generate a distribution of cells with differing levels of
competence\cite{hecht2006cpt}. A small subset of the cells at any given
time end up being receptive to this information exchange, whereas the
rest remain closed. The competent subset changes with time, so
eventually, all cells will at some time be able to accept foreign
genetic material. This competence mechanism is then the microbial analogue of `trust'
in our model. This dynamic may be analogous to the leader phase in
our model. Here, the information amplification takes place when a
subset of cells exchange material and either live or die as a result.
The surviving exchanges are then passed on to the local population,
amplifying the induced information. An analogous process to `taxation' (e.g. resource subsidy)
may occur via a form of symbiosis or biofilm formation, such that nutrient resources are shared.

\section{Conclusions}

We have shown that a model of communicating agents that divide their
time between information generation and information usage has three
distinct phases of organization corresponding to structures
identifiable in human political systems. The flow of information
between agents in the system is critical to this phase structure. If
agents can exchange resources in a way that does not permit cheating,
then the optimal structure is to have a small number of leaders that
scales logarithmically with the system size, and a larger number of
workers. Fluctuations in the reliability of agents tend to emphasize
the communal phase over the leader phase.

The phase transitions predicted by this model are all first order in
nature. As such, in a situation in which the agents are approaching
equilibrium dynamically, the various phases can coexist over much of
the parameter space. This makes sense when one looks at the diversity
of actual political systems in existence, on both the local and
national scales. The transition to the leader phase from a communal
phase takes the form of an inhomogeneous decay in the levels of
decision making of the agents in the system, leaving one agent in
charge by default. In a dynamical version of this model in which the
distribution of agents changes with time, the transition between
different leader agents could be studied.

This model has a relatively simple phase structure, as only the
thinking value and trust levels are allowed to vary. The addition of
spatial considerations, information exchange costs, lying, resource
exchange with cheating, or other such factors could vastly increase the
diversity of phases exhibited by the model.

\section{Acknowledgements}

We would like to thank Ira Carmen for discussion on political systems
and his interest in this work, and Tom Butler for useful discussions on
the mathematics of the multiple-option connected phase.  We acknowledge
partial support from the National Science Foundation through grant
number NSF-EF-0526747.  Nicholas Guttenberg was partially supported by
the University of Illinois Distinguished Fellowship.

\bibliographystyle{apsrev}
\bibliography{politics}

\begin{figure}[t]
\includegraphics[width=\columnwidth,angle=0]{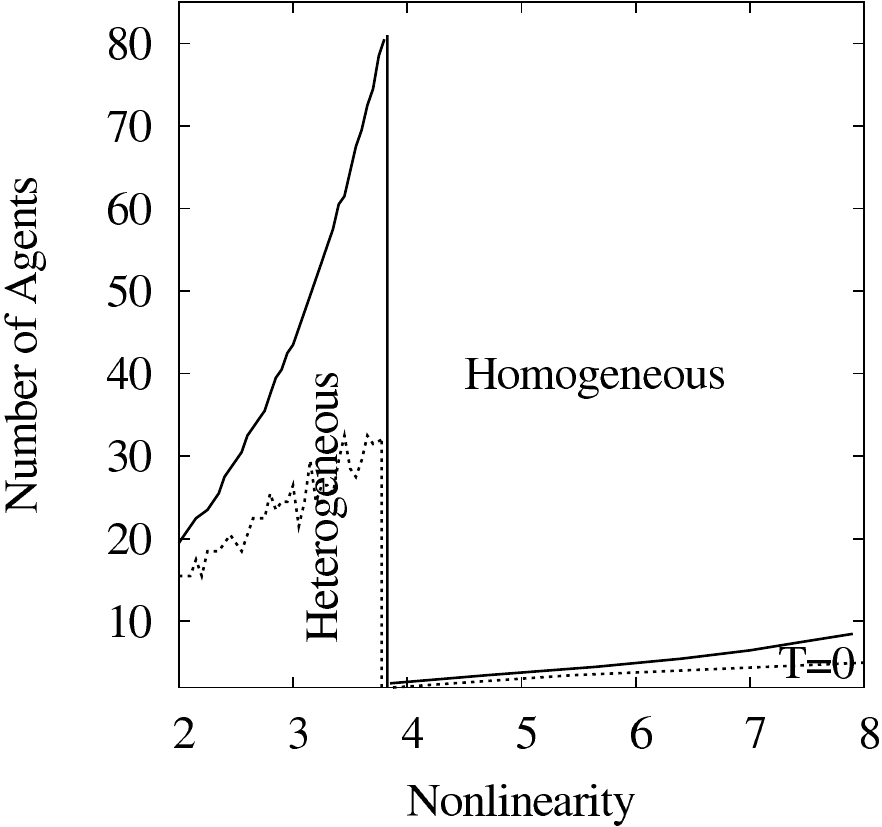}
\caption{Phase diagram for $O=10$ in the space of the nonlinearity
$\alpha$ and number of agents $N$. The phase transition from the
heterogeneous phase as $N$ increases is due to the communal phase being
more efficient than a selfish leader phase. The phase transition as
$\alpha$ increases is due to the transition of the isolated phase to a
$T=0$ phase. The dotted lines show the phase boundaries when Gaussian
fluctuations with a standard deviation of $0.1$ are added to the $T$
value of each agent. } \label{PhaseDiag1}
\end{figure}

\begin{figure}[t]
\includegraphics[width=\columnwidth,angle=0]{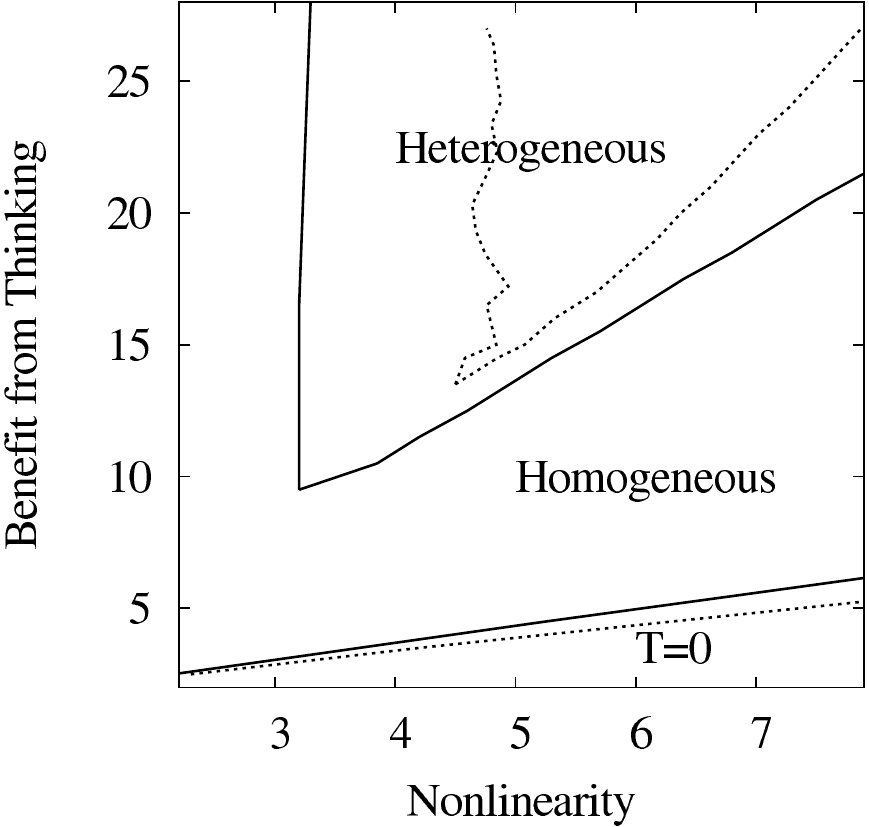}
\caption{Phase diagram for $N=50$ in the space of the nonlinearity
$\alpha$ and thinking benefit $O$. The dotted lines show the phase
boundaries when Gaussian fluctuations with a standard deviation of
$0.1$ are added to the $T$ value of each agent. } \label{PhaseDiag2}
\end{figure}

\end{document}